The Evaluation of Breathing 5:5 effect on resilience, stress and balance center measured by Single-Channel EEG

**Authors:**
Eliezer Yahalom[1]*, Neta Maimon[2] Lior Molcho[1], Talya Zeimer[1], Ofir Chibotero[1], Nathan Intrator[1,3]
[1] Brain in Heart™
[2] Neurosteer Inc, NYC
[3] Tel Aviv University
*Corresponding author

**Abstract**

**Background:** Slow-paced breathing is a promising intervention for reducing anxiety and enhancing emotional regulation through its effects on autonomic and central nervous system function. This study examined the neurophysiological and subjective effects of a 5:5 breathing protocol on stress-related EEG biomarkers using a mobile single-channel EEG system.

**Methods:** Thirty-eight healthy adults were randomly assigned to either an intervention group (n = 20), which completed two sessions spaced two weeks apart and daily practice of 5 minutes in the morning and 5 at night special Brain in Heart connection breathing, or a control group (n = 18), which completed one session. In each session, participants underwent an auditory EEG assessment with resting, mental load, and startle conditions. The intervention group completed a guided breathing session during the first visit and practiced the technique between sessions. EEG biomarkers (ST4, Alpha, Delta, Gamma, VC0) and subjective anxiety levels (STAI) were assessed pre- and post-intervention.

**Results:** A significant reduction in Gamma power was observed in the intervention group immediately following the first breathing session during mental load (p = .002), indicating acute stress reduction. Across sessions, long-term breathing practice led to increased Alpha and Delta power and reduced ST4 activity, suggesting cumulative improvements in emotional regulation and cognitive efficiency. Correlational analyses revealed that changes in VC0 and Alpha were significantly associated with subjective reports of tension, focus difficulty, and calmness.

**Conclusion:** Guided slow-paced breathing at a 5:5 rhythm produces both immediate and sustained effects on neural markers of stress and cognition, with corresponding improvements in subjective anxiety. These findings support the use of EEG-based monitoring to evaluate breath-based interventions and suggest a scalable approach for real-time emotional self-regulation1.

1. Introduction

Breathing is one of the few autonomic functions that can be consciously regulated, providing a direct pathway for influencing physiological and psychological states. In recent decades, controlled breathing techniques have been increasingly recognized as potent tools for emotion regulation, anxiety reduction, and stress resilience. One particularly effective method is slow-paced breathing at a frequency of approximately 0.1 Hz, commonly operationalized as a 5:5 rhythm—five seconds of inhalation followed by five seconds of exhalation. This rhythm has been shown to synchronize respiratory and cardiovascular oscillations, optimize baroreflex sensitivity, and enhance heart rate variability (HRV), all of which contribute to a parasympathetic-dominant state (Lehrer & Gevirtz, 2014; Zaccaro et al., 2018). These physiological responses are closely linked to improvements in subjective well-being and reductions in both state and trait anxiety (Perciavalle et al., 2017; McCraty & Zayas, 2014).

Beyond its autonomic effects, slow breathing has also been proposed to impact central nervous system structures involved in mood, cognition, and consciousness. The pineal gland, located near the center of the brain, is traditionally recognized for its role in circadian rhythm regulation via melatonin secretion (Arendt, 1995). Recent literature suggests the pineal gland may also be involved in altered states of consciousness, meditation, and emotional regulation (Muehsam et al., 2021). Melatonin, the principal hormone secreted by the pineal gland, is known for its anxiolytic and neuroprotective effects (Reiter et al., 2007). Experimental studies have shown that deep breathing or meditation can elevate melatonin levels and modulate cortical oscillations, suggesting a pathway through which breath control may influence emotional state (Harinath et al., 2004)[1].

Despite growing interest in these mechanisms, relatively few studies have empirically linked structured breathing interventions to real-time neural changes during cognitive and emotional tasks. The current study aimed to address this gap by evaluating whether a guided 5:5 breathing protocol could induce measurable changes in neurophysiological markers of stress, as well as reductions in self-reported anxiety. We employed a validated auditory cognitive task battery containing both high mental load and acute startle conditions, in combination with a wearable single-channel EEG device[1].

We used the Neurosteer high-density resolution EEG (hdrEEG) system, a compact, mobile system composed of a three-electrode forehead sensor and cloud-based analytics platform. Unlike traditional multi-electrode EEG systems, the Neurosteer device employs proprietary signal processing algorithms, including harmonic analysis and machine learning, to extract real-time biomarkers from single-channel data. These biomarkers have been validated in numerous cognitive and clinical contexts[1].

In this study, we focused on five EEG features: ST4, Alpha, Delta, Gamma, and VC0. ST4 is a machine-learning-derived feature that has shown sensitivity to cognitive load and individual performance in working memory tasks (Maimon et al., 2022; Molcho et al., 2023). Alpha oscillations are well-documented indicators of relaxed, internally directed attention, and their amplitude has been inversely related to perceived stress and anxiety (Klimesch, 1999; Katmah et al., 2021). Delta activity, while most prominent during sleep, is also implicated in emotional processing and has been observed to increase during meditative states or following relaxation protocols (Başar et al., 2001). Gamma oscillations are typically associated with integrative cognitive processes and working memory; however, elevated gamma activity under stress has been linked to hyperarousal, while reductions may signal successful downregulation following relaxation (Gärtner et al., 2014). VC0, another proprietary marker, was identified in previous studies involving auditory n-back tasks, where it served as a sensitive index of cognitive load and attentional control (Maimon et al., 2022). Given its robustness in distinguishing cognitive states, VC0 was also included as a candidate marker for stress reduction[1].

The experimental design consisted of two groups. The intervention group (n = 18) participated in two lab visits, spaced two weeks apart. During each visit, participants completed an initial auditory EEG assessment, which included three task conditions: resting state, high mental load, and startle. These conditions were designed to elicit varying levels of cognitive demand and stress, enabling us to examine both baseline and task-evoked EEG patterns. Next, the intervention group subsequently underwent a 20-minute video guided breathing session. This session followed a structured 5:5 rhythm and included verbal and auditory cues designed to enhance focus, promote emotional downregulation, and support internal attention toward interoceptive sensations. The auditory component featured calming alpha and theta frequency tones, along with a 963 Hz tone commonly linked to pineal gland activation. Verbal instructions, recorded from a TV screen, were designed to encourage internal focus and self-regulation. The protocol aimed to induce relaxation by integrating breath entrainment with auditory stimulation, targeting both physiological and psychological stress pathways. Instructions emphasized central awareness, referencing inner rhythm and equilibrium, and drawing upon the hypothesized role of the pineal gland in mediating breath-induced neurochemical states. Immediately following the breathing session, participants repeated the auditory EEG assessment1.

After the first session, participants in the intervention group were encouraged to continue practicing the tactic 5:5 breathing technique daily in their natural environment. Approximately two weeks later, they returned for a second session that followed the same structure. This allowed us to examine both acute effects (immediate changes from pre- to post-breathing in session one) and cumulative effects (overall differences between session one and session two following continued practice). Based on the literature, we hypothesized that participants would show significant increases in Alpha and Delta power, and decreases in ST4 and Gamma activity, immediately following the breathing session. We also expected a sustained modulation of VC0, reflecting reduced mental load and stress reactivity across sessions. In contrast, we anticipated weaker or absent immediate effects in the second session, due to baseline shifts in arousal or ceiling effects in regulatory adaptation1.

A control group (n = 18) completed a single session during which they watched a calming nature film for 20 minutes between two identical auditory EEG assessments. This passive control condition allowed us to differentiate the effects of guided breathing from general relaxation or task repetition1.

To complement the neurophysiological data, all participants completed the State–Trait Anxiety Inventory (STAI; Spielberger et al., 1983) before and after each session. The STAI distinguishes between transient (state) anxiety and stable anxiety dispositions (trait). In this study, we focused on the pre–post changes in state anxiety scores, particularly in the first session, to assess whether subjective reductions in anxiety corresponded with shifts in EEG biomarkers. We hypothesized that participants in the intervention group would report larger reductions in state anxiety compared to the control group and that these changes would correlate with biomarker alterations, especially in Alpha, Delta, and ST41.

Taken together, this study aimed to evaluate whether a single-channel EEG system could detect meaningful neural changes associated with a brief, structured breathing intervention. By integrating auditory cognitive assessments, proprietary EEG biomarkers, and subjective anxiety measures, we sought to provide a scalable model for real-time stress detection and regulation. If successful, this approach could support the development of accessible interventions for emotional regulation in both clinical and high-performance contexts1.

2. Methods

2.1 Participants
Ethical approval for the study was granted by the Ethics Committee of Tel Aviv University on May 2, 2024. All participants provided written informed consent prior to participating in the experiment. Thirty-eight healthy adult volunteers participated in the study (28 females, 10 males; mean age = 54.34 years, SD = 14.11). All participants were screened to ensure they had no history of neurological or psychiatric disorders. The study was conducted in accordance with the Declaration of Helsinki and approved by the relevant ethics committee1.

2.2 Apparatus
EEG data were recorded using the Neurosteer® single-channel high dynamic range EEG (hdrEEG) Recorder. A three-electrode medical-grade patch was applied to each participant's forehead using dry gel for optimal signal transduction. The monopolar electrode configuration included electrodes at Fp1 and Fp2, with a reference at Fpz (International 10/20 system). The EEG signal was sampled at 500 Hz and transmitted wirelessly for real-time data processing. EEG signals were processed using time-frequency analysis to extract neural features. Biomarkers A0, ST4, VC0, T2, and VC12 were derived using Neurosteer's proprietary machine-learning algorithms. Power spectral density was computed using fast Fourier transform to analyze delta (0.5–4 Hz), theta (4–7 Hz), alpha (8–15 Hz), beta (16–31 Hz), and gamma (32–45 Hz) frequency bands.

### 2.3 Procedure

Participants were randomly assigned to either an intervention group (two guided breathing sessions, 2 weeks apart) or a control group (single session). All sessions were conducted in a controlled lab environment. At the beginning of each session, participants completed the State–Trait Anxiety Inventory (STAI) and additional questions on stress perception, sleep quality, and current emotional state. Following the questionnaire, participants underwent an initial EEG recording using the Neurosteer hdrEEG system. The assessment included:

- Resting-state EEG with eyes open
- Musical detection task (responding to brief target stimuli)
- Musical n-back task (2-back level, engaging working memory)
- Random startle stimuli (unpredictable 100 dB beeps) to induce stress responses

After the initial EEG and task battery, only the intervention group proceeded to a relaxation environment for a 20-minute guided breathing session ("Tactic Breathing 5:5" by Dr. Eliezer Yahalom: five seconds inhale, five seconds exhale, nasal breathing only, synchronized with auditory guidance). The auditory component included calming alpha and theta frequency tones and a 963 Hz tone associated with pineal gland activation. Verbal instructions recorded from TV screen, promoted internal attention and self-regulation. The protocol aimed to evoke relaxation by combining breath entrainment and auditory stimulation, targeting physiological and psychological stress pathways.

Immediately following the breathing intervention, participants repeated the full EEG and auditory assessment protocol. Approximately 2 weeks later, the intervention group returned for a second session following the same structure as the first.

The control group, in contrast, completed only a single session. After the initial EEG and auditory task battery, control participants watched a calming nature film for 20 minutes in a comfortable room. No breathing intervention or specific instructions were provided. After the film, they completed the same EEG protocol and post-assessment as the intervention group. This control condition allowed for distinguishing the specific effects of guided breathing from general relaxation or task habituation.

### 2.4 Statistical Analysis

EEG data were averaged across trials for each participant and categorized into three task conditions: resting state, startle, and mental load. The resting state condition included all resting state tasks, the startle condition included the abrupt auditory beeps, and the mental load condition combined both the n-back and detection tasks. For each participant, neural activity within these categories was averaged to create a single value per feature and condition.

The primary analysis focused on comparing pre- and post-intervention EEG responses for each condition and session, evaluating changes in neural biomarkers (ST4, VC0) as well as frequency bands (alpha, delta and gamma). First an LMM with session, pre post and condition were conducted to see the main effect. For significant biomarkers, paired t-tests were conducted to assess differences between pre- and post-breathing intervention within each session and condition and to compare neural activity between conditions (e.g., rest vs. mental load) within the pre or post session. A Benjamini-Hochberg false discovery rate (FDR) correction was applied to control for multiple comparisons.

In addition to these comparisons, Pearson correlations were calculated between the difference in activity between pre and post breathing session and the subjective questionnaire responses. This analysis aimed to identify associations between neural features and subjective reports of stress and anxiety, with a specific focus on composite questionnaire measures such as momentary tension and lack of focus (aggregating relevant items). This approach allowed us to explore how neural markers of cognitive and stress-related activity correspond to subjective experience, both at baseline and following the relaxation intervention.

3. Results

**3.1 Pre vs. post breathing effects and sessions**

In the LMM models Gamma showed a significant main effect the short 20 minutes breathing session ($p = 0.008$, see appendix for all LMM results). Post hoc analysis revealed that for the manipulation group, in the first session there was a significant decrease in Gamma power between pre and post 20 minutes breathing intervention during the mental load condition ($p = 0.002$). No other differences between pre and post 20 minutes breathing intervention were found in the second session.

The difference between the two sessions of the intervention across the pre and post 20 minutes of breathing was significant for ST4: the resting state condition was lower after 2 weeks of breathing exercise ($p = 0.014$), and for Delta and Alpha power during the mental load condition ($p = 0.011$ and $p = 0.027$ respectively). For means and SE of the intervention group see Figure 1, and for control group see Figure 2.

For the control group (nature movie), none of the differences between pre and post were significant.

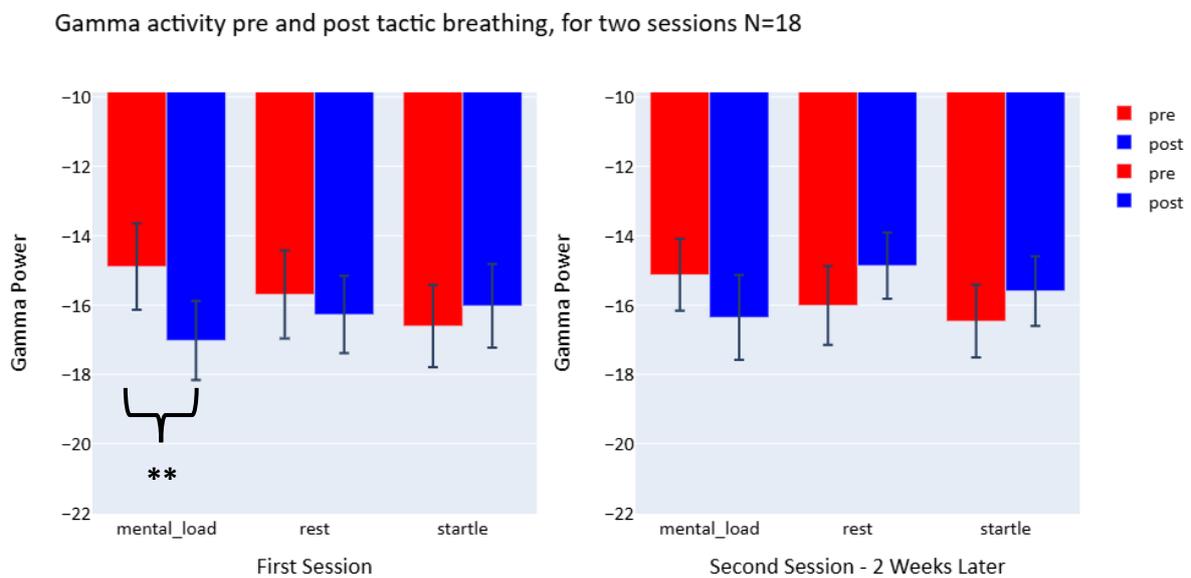

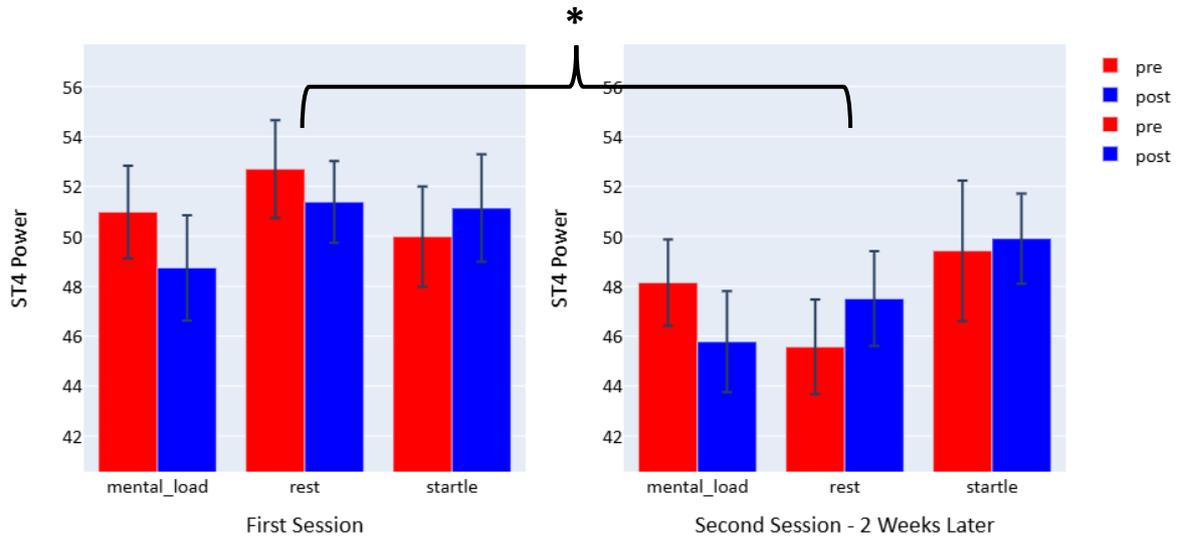

ST4 activity pre and post tactic breathing, for two sessions N=18

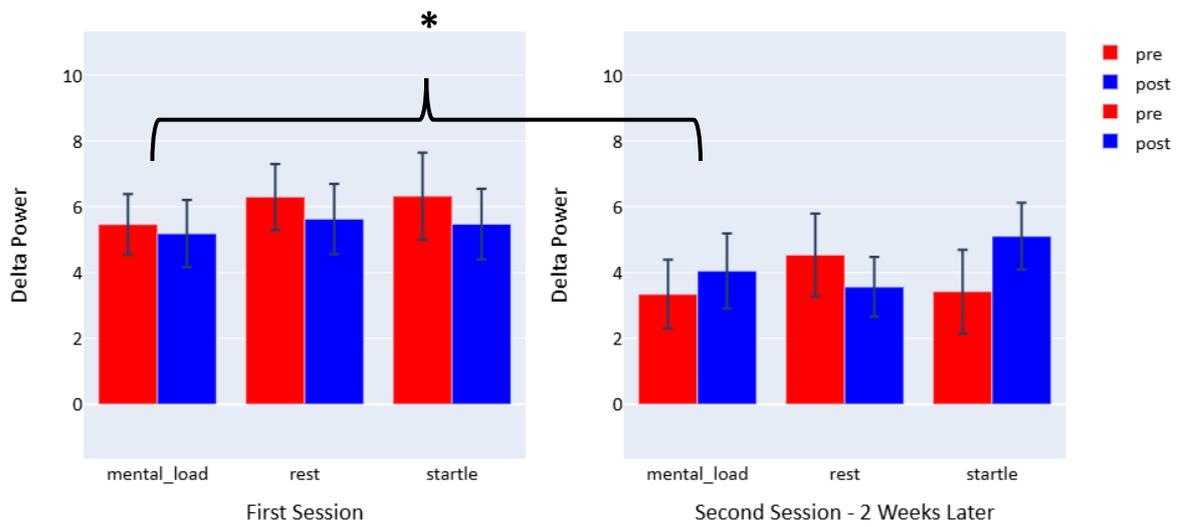

Delta activity pre and post tactic breathing, for two sessions N=18

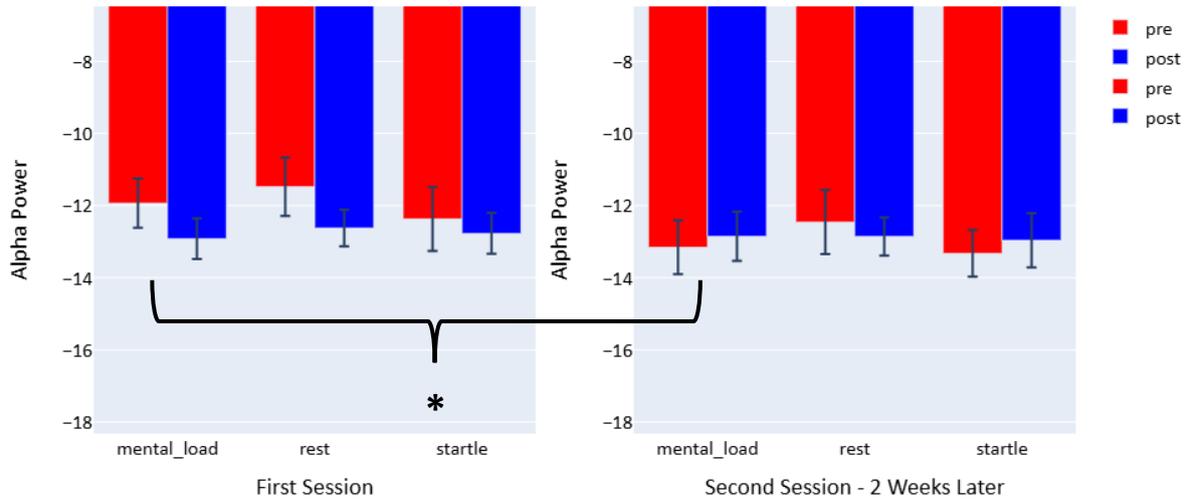

**Figure 1A:** ST4, Delta, Alpha and Gamma biomarker normalized 1-100 for the intervention group by cognitive load condition (mental load, rest, startle) before (blue) and after (red) the breathing-based intervention for first session (left) and second session (right). Error bars represent standard errors of the mean (SEM). * indicates p < 0.05, ** p < 0.01, *** p < 0.001.

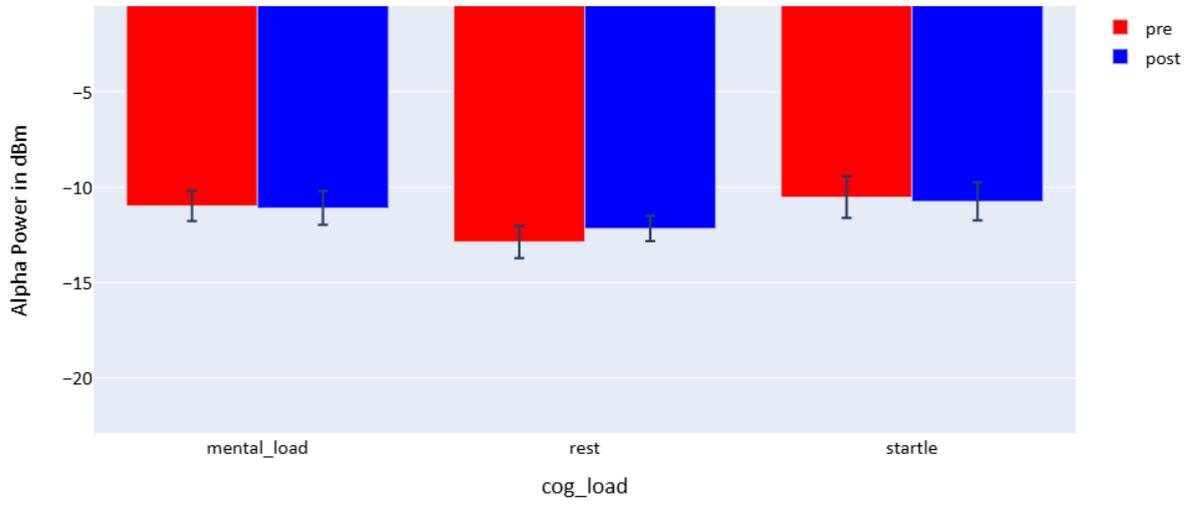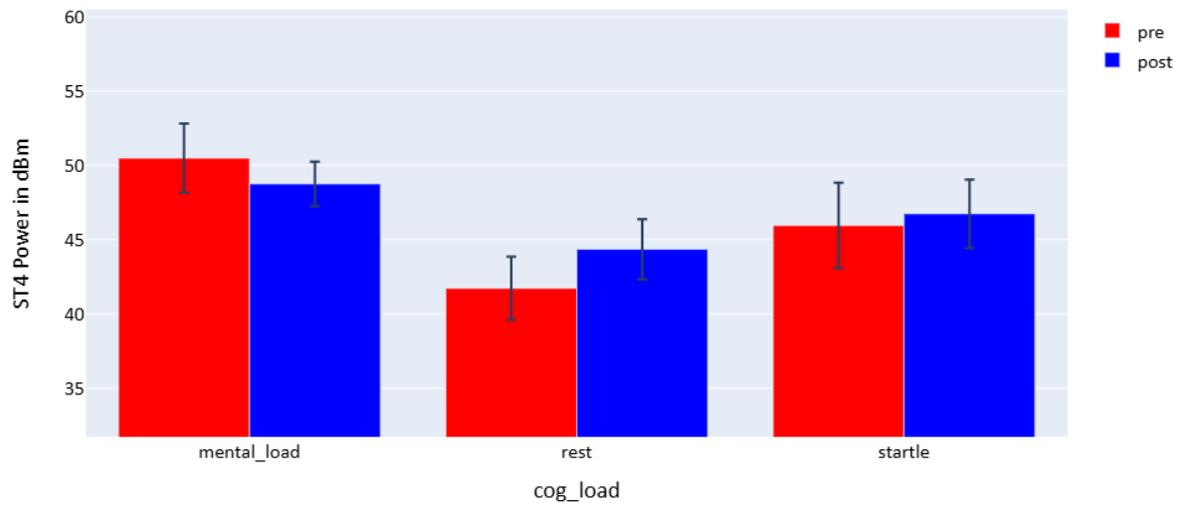

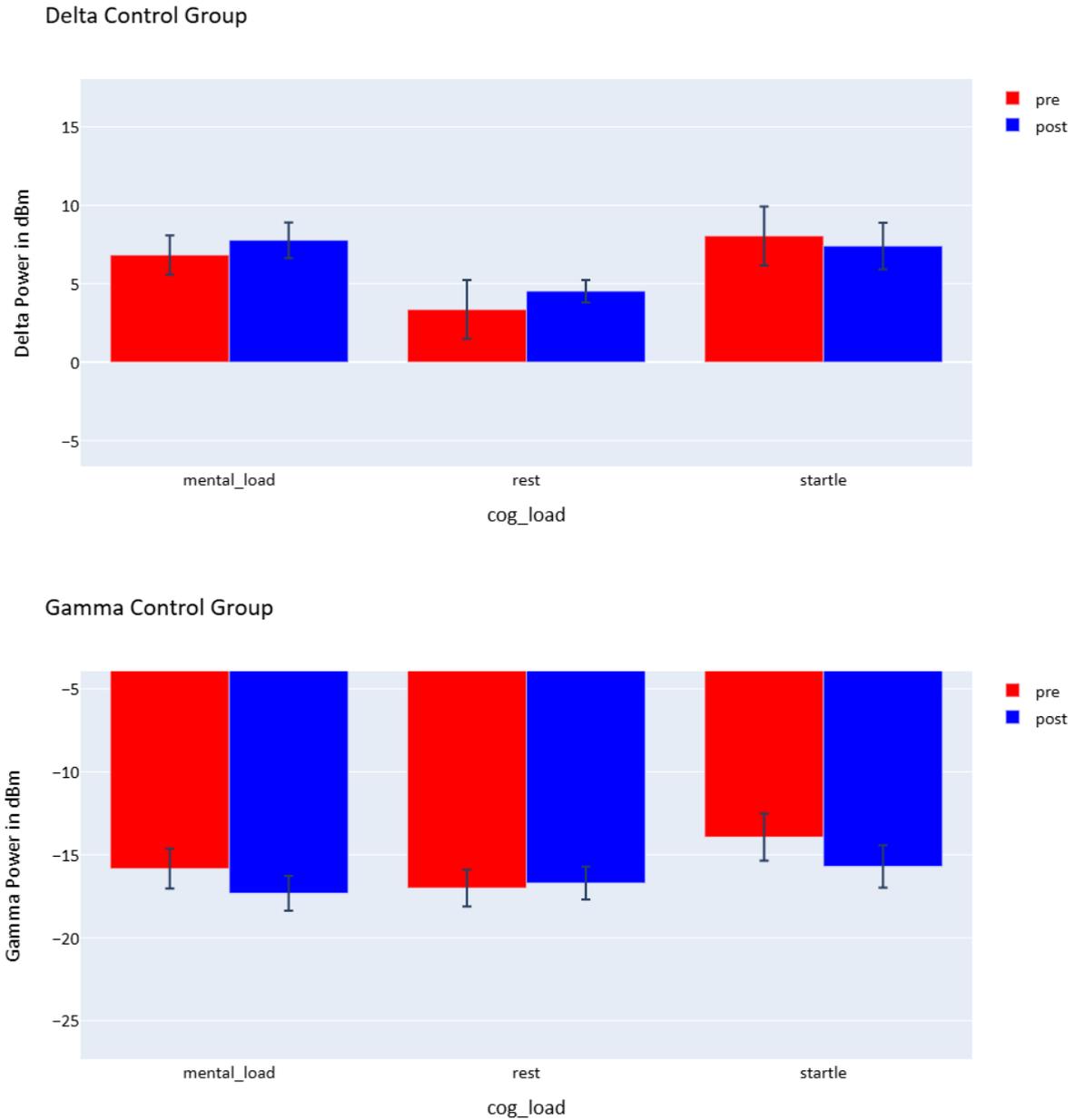

**Figure 1B:** ST4, Delta, Alpha and Gamma for the control group by cognitive load condition (mental load, rest, startle) before (blue) and after (red) the breathing-based intervention. Error bars represent standard errors of the mean (SEM). No differences were found significant.

### 3.3 Correlations

**Correlations between difference of hdfEEG activity between pre and post breathing session and subjective reports**

To further explore the effect of the breathing-based stress reduction intervention on neural activity, we calculated the difference between pre- and post-intervention EEG responses for each feature and condition. We then examined whether these changes in neural activity were associated with subjective reports of stress, anxiety, and focus by correlating the feature-specific pre-post differences with questionnaire responses.

VC0 difference (VC0_diff) was positively correlated with momentary tension (now_tens), r = 0.52, p = 0.005, lack of focus (now_no_focus), r = 0.41, p = 0.029, and fear (now_fear), r = 0.38, p = 0.047, and negatively correlated with perceived control (now_control), r = -0.39, p = 0.043. VC0_diff also showed a strong positive correlation with the combined momentary tension and focus difficulty index (now_tens_no_focus), r = 0.56, p = 0.002, indicating that participants with larger VC0 neural changes after the intervention reported higher momentary tension and difficulty focusing.

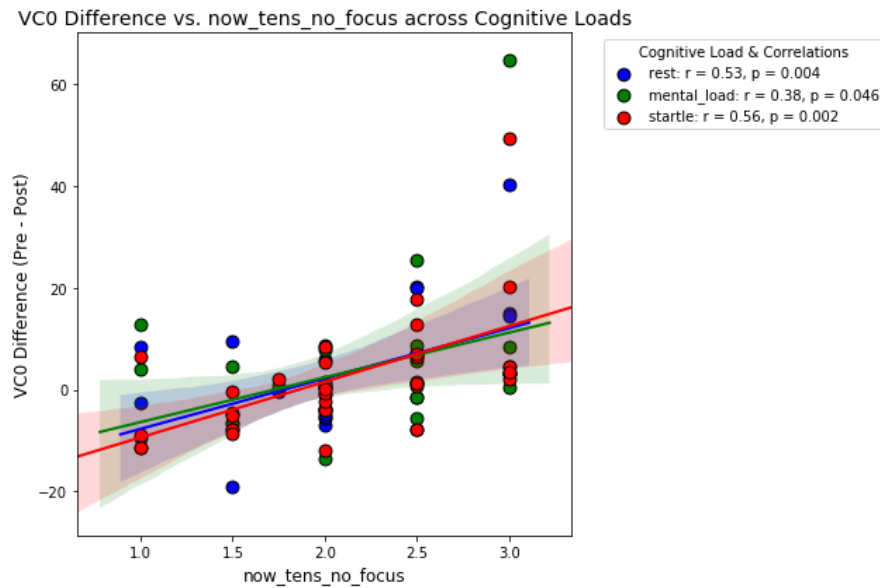

**Figure 4:** Correlation between VC0 difference (pre-post breathing intervention) and the combined momentary tension and focus difficulty index (now_tens_no_focus) across resting state (blue), mental load (green), and startle (red) conditions. Larger VC0 changes were associated with higher momentary tension and difficulty focusing.

Another consistent correlation was found between alpha difference (Alpha_diff) and sleep hours. It was negatively correlated with sleep hours, during the startle $r = -0.42$, $p = 0.027$, the mental load $r = -0.04$, $p = .037$, (see Figure 5) suggesting that participants who slept less showed greater changes in alpha activity in response to the startle following the intervention. Finally, alpha difference (Alpha_diff) was also positively correlated with general calmness (gen_calm), $r = 0.53$, $p = 0.004$.

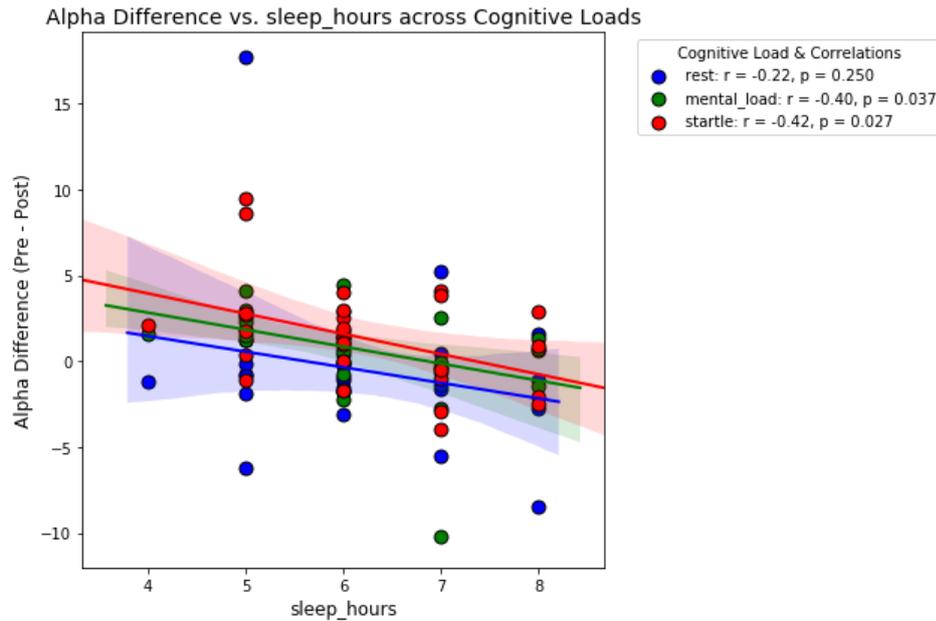

**Figure 5:** Correlation between alpha power difference (pre-post breathing intervention) and sleep hours across resting state (blue), mental load (green), and startle (red) conditions. Greater changes in alpha activity were associated with fewer reported sleep hours.

Additionally, ST4 difference (ST4_diff) was positively correlated with general anxiety (gen_anx), $r = 0.41$, $p = 0.032$, and with general calmness (gen_calm), $r = 0.46$, $p = 0.015$, suggesting a complex relationship between ST4 neural dynamics and subjective stress-calmness balance.

### 4. Discussion

The present study investigated the neurophysiological and subjective effects of a slow-paced 5:5 breathing intervention on stress-related brain activity using single-channel EEG and a task battery incorporating mental load, rest, and startle conditions. Our findings provide converging evidence that a short breathing intervention can induce immediate neural changes—particularly in Gamma activity—and that continued practice over 2 weeks leads to sustained modulations in ST4, Alpha, and Delta biomarkers. Moreover, these neural changes were meaningfully associated with self-reported stress, anxiety, focus, and calmness, reinforcing the potential of structured breathwork as a reliable stress-regulation strategy.

We first examined the immediate effects of a single 20-minute guided breathing session. A significant reduction in Gamma power was observed in the intervention group during the mental load condition after the first session. This effect was absent in both the second session and the control group, suggesting that the Gamma reduction reflects an acute neural response to the breathing intervention rather than task repetition or general relaxation. Elevated Gamma activity has previously been associated with cognitive overload, hyperarousal, and stress-related attentional dysregulation (Gärtner, Grimm, & Bajbouj, 2014; Başar et al., 2001). Therefore, the observed decrease in Gamma power may signal a shift toward a more regulated and less reactive cognitive state, particularly under demanding conditions. This aligns with prior evidence that slow breathing induces parasympathetic dominance, reduces noradrenergic tone, and supports cortical inhibition (Lehrer & Gevirtz, 2014; Zaccaro et al., 2018).

Importantly, we also found longer-term changes in neural biomarkers after continued breathing practice over 2 weeks. Compared to the first session, ST4 values during resting state were significantly lower in the second session. ST4 is a machine learning-derived biomarker known to track cognitive effort and performance in auditory working memory tasks (Maimon, Molcho, & Intrator, 2022; Molcho, Maimon, & Intrator, 2023). The reduction in resting ST4 suggests a more efficient or less burdened baseline cognitive state, possibly reflecting decreased anticipatory arousal or improved regulation following the breathing practice. This pattern supports the hypothesis that slow, regular breathing entrains autonomic-cortical loops and enhances the brain's capacity to maintain a calm resting baseline, consistent with theoretical models of brain-heart coherence (Muehsam et al., 2021).

In addition, we found increased Alpha and Delta power during mental load in the second session compared to the first. Alpha rhythms are typically enhanced during internal focus and relaxation and suppressed under anxiety or task-induced stress (Klimesch, 1999; Katmah et al., 2021). Increased Alpha during a cognitively demanding condition may therefore reflect improved emotional buffering and attentional control. Similarly, Delta oscillations, although primarily associated with deep sleep, also play a role in affective modulation and interoceptive awareness during waking states (Başar et al., 2001). The observed Delta increase may index a greater engagement of neural systems that support emotional resilience, possibly facilitated by repeated breath-induced parasympathetic activation and its downstream cortical effects (Zaccaro et al., 2018).

The correlational analyses further supported the relevance of these neural modulations. VC0, an EEG feature previously validated in auditory working memory paradigms, was positively associated with self-reported momentary tension, lack of focus, and fear, and negatively associated with perceived control. Participants who exhibited greater neural changes in VC0 following the breathing session also reported higher baseline difficulty with stress regulation. This suggests that VC0 may be particularly sensitive to momentary psychological states and could serve as a dynamic marker of emotional reactivity (Maimon et al., 2022).

Notably, Alpha difference scores were negatively correlated with sleep hours and positively correlated with general calmness. This implies that participants who were more physiologically vulnerable (e.g., due to poor sleep) showed stronger Alpha modulation in response to breathing, potentially reflecting a compensatory neuroregulatory mechanism. ST4 changes also correlated with both general anxiety and calmness, highlighting the complexity of this feature's role in mediating cognitive-affective balance.

Together, these results provide evidence for both immediate and cumulative effects of slow-paced breathing on neural activity and subjective well-being. The short-term Gamma reduction may reflect decreased stress reactivity and improved top-down control, while the long-term increases in Alpha and Delta, and reductions in ST4, suggest sustained improvements in baseline emotional regulation and cognitive efficiency.

From a mechanistic perspective, these effects are likely mediated by vagal nerve activation, baroreflex sensitivity enhancement, and downstream modulation of cortical rhythms associated with emotional and cognitive processing (Lehrer & Gevirtz, 2014). The inclusion of auditory guidance targeting specific frequency bands—including alpha, theta, and the 963 Hz tone often referred to in pineal activation protocols—may have amplified these effects by synchronizing auditory entrainment with interoceptive focus. While the pineal gland has historically been associated with circadian rhythms through melatonin regulation (Arendt, 1995), recent work suggests that rhythmic input—such as through breath and frequency—may modulate pineal-related neuromodulatory states (Harinath et al., 2004; Muehsam et al., 2021). Although speculative, this framework helps explain how structured breathing, especially when directed with intention and frequency-specific guidance, could alter both subjective consciousness and measurable EEG patterns.

This study offers several important contributions. First, it demonstrates that a single-channel EEG system can detect meaningful neural changes related to stress and relaxation, even with minimal sensor setup. Second, it shows that breath-based interventions, when structured and sustained, can modify both cortical activity and subjective anxiety. Third, it highlights specific EEG biomarkers—Gamma, ST4, Alpha, Delta, and VC0—as sensitive indicators of these effects, paving the way for scalable mental health monitoring.

However, several limitations should be acknowledged. The sample size, while adequate for detecting medium effects, may limit generalizability. Future studies should expand to clinical populations, explore dose-response effects, and incorporate physiological measures such as heart rate variability or salivary cortisol to deepen mechanistic understanding. The control condition, though providing a baseline for general relaxation, did not include a sham breathing protocol, and future designs could compare different types of breathwork.

In conclusion, our findings support the hypothesis that guided slow-paced breathing at a 5:5 rhythm produces measurable changes in EEG biomarkers associated with stress and cognitive load. Both immediate and sustained effects were observed, corresponding with improvements in subjective stress and anxiety. These results reinforce the growing literature on breath-based neuroregulation and provide a foundation for the development of real-time, brain-based feedback systems for emotional self-regulation.

5. **Declaration of generative AI and AI-assisted technologies in the writing process**

During the preparation of this work the author(s) used chatGPT in order to to improve the readability and language of the manuscript. After using this tool/service, the authors reviewed and edited the content as needed and take full responsibility for the content of the published article.

6. **Declaration of interests**

Neta B. Maimon reports equipment was provided by Neurosteer Inc. Neta B. Maimon reports a relationship with Neurosteer Inc that includes: employment. Lior Molcho reports a relationship with Neurosteer Inc that includes: employment. Talya Zeimer reports a relationship with Neurosteer Inc. that includes: employment. Ofir Chibotero reports a relationship with Neurosteer Inc that includes: employment. Nathan Intrator reports a relationship with Neurosteer Inc. that includes: employment and equity or stocks. If there are other authors, they declare that they have no known competing financial interests or personal relationships that could have appeared to influence the work reported in this paper.